\documentclass[preprint,showpacs,superscriptaddress,preprintnumbers,amsmath,amssymb]{revtex4-1}

\usepackage{amsfonts}
\usepackage{mathrsfs}
\usepackage{amsmath}% needed for subequations
\usepackage{color}
\usepackage{graphicx}
\usepackage{bm}% bold maths
\usepackage{amssymb}
\usepackage{xspace}
\usepackage{epstopdf}
\usepackage{dcolumn}% Align table columns on decimal point
\usepackage{longtable}
\usepackage{multirow}
\usepackage[colorlinks=true, letterpaper=true, pdfstartview=FitV, linkcolor=blue, citecolor=blue, urlcolor=blue]{hyperref}

\usepackage{ulem}

\begin{document}

%\preprint{APS/123-QED}

\title{The emergent linear Rashba spin-orbit coupling offering the fast manipulation of hole-spin qubits in germanium}
\author{Yang Liu}
\affiliation{State Key Laboratory of Superlattices and Microstructures, Institute of Semiconductors,Chinese Academy of Sciences, Beijing 100083, China}
\affiliation{Center of Materials Science and Optoelectronics Engineering, University of Chinese Academy of Sciences, Beijing 100049, China}

\author{Jia-Xin Xiong}
\affiliation{State Key Laboratory of Superlattices and Microstructures, Institute of Semiconductors,Chinese Academy of Sciences, Beijing 100083, China}
\affiliation{Center of Materials Science and Optoelectronics Engineering, University of Chinese Academy of Sciences, Beijing 100049, China}

\author{Zhi Wang}
\affiliation{State Key Laboratory of Superlattices and Microstructures, Institute of Semiconductors,Chinese Academy of Sciences, Beijing 100083, China}

\author{Wen-Long Ma}
\affiliation{State Key Laboratory of Superlattices and Microstructures, Institute of Semiconductors,Chinese Academy of Sciences, Beijing 100083, China}

\author{Shan Guan}
\email{shan\_guan@semi.ac.cn}
\affiliation{State Key Laboratory of Superlattices and Microstructures, Institute of Semiconductors,Chinese Academy of Sciences, Beijing 100083, China}

\author{Jun-Wei Luo}
\email{jwluo@semi.ac.cn}
\affiliation{State Key Laboratory of Superlattices and Microstructures, Institute of Semiconductors,Chinese Academy of Sciences, Beijing 100083, China}
\affiliation{Center of Materials Science and Optoelectronics Engineering, University of Chinese Academy of Sciences, Beijing 100049, China}
\affiliation{Beijing Academy of Quantum Information Sciences, Beijing 100193, China}

\author{Shu-Shen Li}
\affiliation{State Key Laboratory of Superlattices and Microstructures, Institute of Semiconductors,Chinese Academy of Sciences, Beijing 100083, China}
\affiliation{Center of Materials Science and Optoelectronics Engineering, University of Chinese Academy of Sciences, Beijing 100049, China}

\begin{abstract}
The electric dipole spin resonance (EDSR) combining strong spin-orbit coupling (SOC) and electric-dipole transitions facilitates fast spin control in a scalable way, which is the critical aspect of the rapid progress made recently in germanium (Ge) hole-spin qubits. However, a puzzle is raised because centrosymmetric Ge lacks the Dresselhaus SOC, a key element in the initial proposal of the hole-based EDSR. Here, we demonstrate that the recently uncovered finite \textit{k}-linear Rashba SOC of 2D holes offers fast hole spin control via EDSR with Rabi frequencies in excellent agreement with experimental results over a wide range of driving fields. We also suggest that the Rabi frequency can reach 500 MHz under a higher gate electric field or multiple GHz in a replacement by [110]-oriented wells. These findings bring a deeper understanding for hole-spin qubit manipulation and offer design principles to boost the gate speed.
\end{abstract}

\maketitle
After demonstrating the all-electrical manipulation of a single hole-spin qubit in gate-defined planar quantum dots (QD) in germanium (Ge) quantum wells (QWs)~\cite{hendrickxSingleholeSpinQubit2020}, remarkably rapid progress has been made in increasing the number of coupled qubits--- doubled every year~\cite{hendrickxFastTwoqubitLogic2020,hendrickxFourqubitGermaniumQuantum2021}. These developments leverage the compelling properties of holes in Ge QWs~\cite{hendrickxGatecontrolledQuantumDots2018,scappucciGermaniumQuantumInformation2020,aggarwalEnhancementProximityinducedSuperconductivity2021} such as: suppressed hyperfine interaction with nuclear sites~\cite{itohHighPurityIsotopically1993, tisotopicallyenriched_2012} resulting in much longer spin coherence times~\cite{bulaev2005spin}; free from the valley degeneracy that is a crucial challenge for the use of silicon electrons as qubits~\cite{zhang_genetic_2013}; low hole effective mass that benefits the desired high tunnel rates for coupled qubits~\cite{PhysRevB.100.041304}; and a strong spin-orbit interaction that is an inherent relativistic effect of the heavy atom~\cite{luoRapidTransitionHole2017}.  Among these properties,  the strong spin-orbit interaction is most striking since it allows for electric-dipole spin resonance (EDSR) controlled by alternating electric fields ~\cite{bulaevElectricDipoleSpin2007,PhysRevLett.109.107201,PhysRevB.87.195307}, bringing about faster spin manipulation in a scalable way~\cite{hendrickxFastTwoqubitLogic2020,scappucciGermaniumQuantumInformation2020,hendrickxFourqubitGermaniumQuantum2021} as opposed to magnetically driven electron spin resonance (ESR) used extensively for manipulation of Si electron spin qubits~\cite{xiao_electrical_2004,koppens_driven_2006,veldhorstTwoqubitLogicGate2015,PhysRevB.85.125312,zajac_resonantly_2018}.

Although EDSR mediated by intrinsic spin-orbit coupling (SOC) has been demonstrated experimentally to coherent manipulate hole spins  in planar Ge QDs with driving frequencies exceeding 100 MHz~\cite{hendrickxFastTwoqubitLogic2020,scappucciGermaniumQuantumInformation2020,hendrickxFourqubitGermaniumQuantum2021}, the underlying microscopic mechanism remains ambiguous~\cite{bulaevElectricDipoleSpin2007,borhaniSpinManipulationRelaxation2012,terrazosTheoryHoleSpinQubits2021,philippNaturalHeavy-hole2021}. An applied alternating current (AC) electric field could induce transitions between spin-up and spin-down states of the lowest SOC-hybridized spin doublet if, only if, in which the Fock-Darwin ground state ($n=0$) coupling to the $n=1$ excited states with opposite spin orientations because of electric-dipole transitions ($\Delta n=\pm 1$ and $\Delta s=0$). The \textit{k}-linear Rashba and Dresselhaus SOC usually provide required $\Delta n=\pm 1$ coupling for electrons confined in gate-defined planar QDs~\cite{golovach2006electric}. However, in common sense, these \textit{k}-linear SOC terms are absent in 2D holes~\cite{winkler2003spin,PhysRevLett.113.086601,luoDiscoveryNovelLinearin2010} since they are in the heavy-hole (HH) subbands. The original EDSR proposal for their hole counterparts~\cite{bulaevElectricDipoleSpin2007,PhysRevLett.109.107201,PhysRevB.87.195307} thus has to rely on the \textit{k}-cubic Dresselhaus SOC, considering it can also couple the ground HH  ($n=0$) to the excited HH $n=1$ states as a result of in-plane wavevector quantization in planar QDs. Unfortunately, such inversion-asymmetry-induced Dresselhaus SOC is nonexistent in centrosymmetric solids, including Si and Ge. To explore efficient EDSR manipulation of hole spins confined in planar Ge QDs Terrazos \textit{et al.} ~\cite{terrazosTheoryHoleSpinQubits2021} resorted to a previously ignored cubic-symmetric component of the \textit{k}-cubic Rashba SOC of 2D holes with a perpendicular magnetic field instead of an in-plane field utilized in experiments~\cite{hendrickxFastTwoqubitLogic2020,hendrickxFourqubitGermaniumQuantum2021}. Nevertheless, such cubic-symmetric Rashba term is far less sizable, if not vanishing, in Ge~\cite{winkler2003spin,philippNaturalHeavy-hole2021}. Therefore, the experimentally achieved rapid EDSR is not yet understood.

\begin{figure}[tb]                
	\centering 
	\includegraphics[width=0.8\columnwidth]{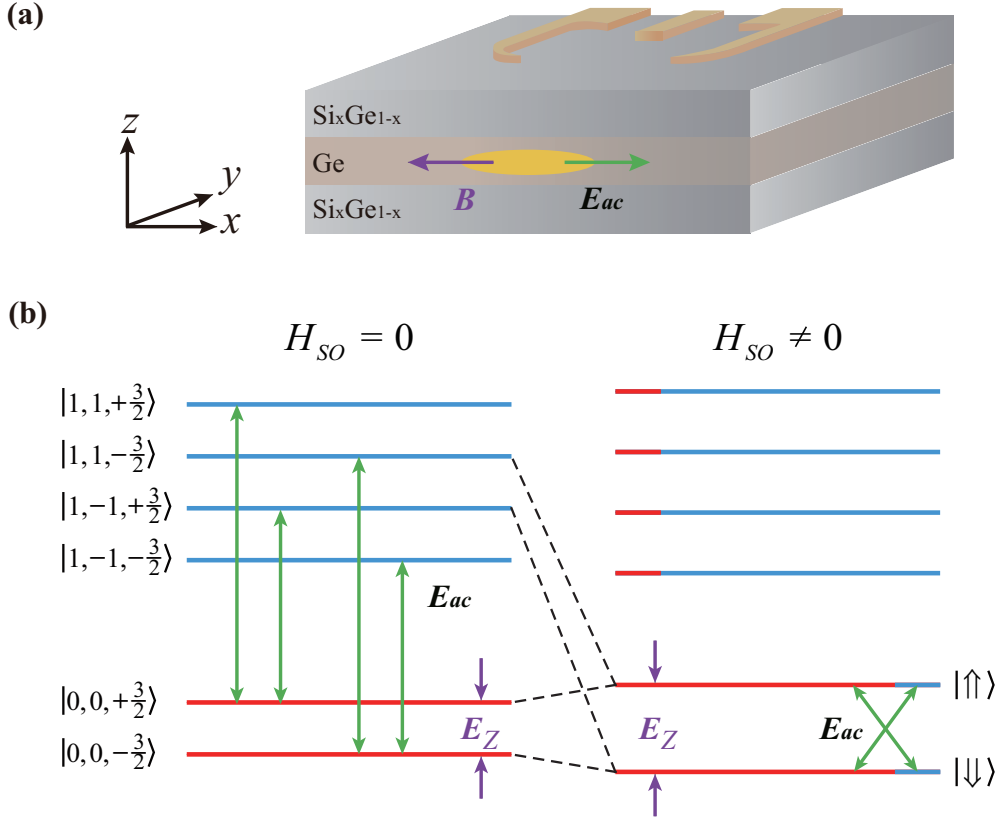} 
	\caption{(a) Schematic illustration of hole state (red) occupying a planar QD, defined by voltage-biased gates on top of the Ge/SiGe QW. The magnetic field $\bf B$ and ac electric field $\bf E_{ac}$  exert on the hole state for EDSR control.  (b) The \textit{k}-linear Rashba SOC mixes the HH ground $n=0$ states with the HH excited $n=1$ states, thus making the EDSR possible.  Rabi oscillations can be achieved if the in-plane ac electric field $\bf E_{ac}$ frequency resonates with the Zeeman splitting $E_Z$ of the lowest SOC-hybridized spin doublet.} 
	\label{fig1}
\end{figure}

This standing puzzle may be resolved by our recently uncovered finite \textit{k}-linear Rashba SOC of 2D holes in, e.g., Ge/Si QWs originating from a combination of local interface-induced HH-light-hole (LH) mixing and direct dipolar intersubband coupling to the external electric field~\cite{xiongEmergenceStrongTunable2021}. Here,  relying on this emergent \textit{k}-linear Rashba SOC supplying the required coupling between $\Delta n=1$ HH states, we develop the EDSR technique for planar QD confined hole spins following Ref.~\onlinecite{bulaevElectricDipoleSpin2007}. Using  a set of experimental device parameters with input Rashba parameter obtained from the atomistic pseudopotential method calculation without  {\it ad hoc} assumptions, we predict a 100 MHz Rabi frequency in excellent agreement with the experimental result of 108 MHz~\cite{hendrickxFastTwoqubitLogic2020}. We also reproduce the experimentally measured electric-field-dependences of Rabi frequency under two investigated magnetic fields of 0.5 and 1.65 T. Consequently, we have solved the puzzle by identifying the newly found linear Rashba SOC contributing to EDSR for rapid control of hole spins confined in planar Ge QDs.

Fig.~\ref{fig1}(a)  schematically shows the experimental setup of gate-defined planar Ge QD formed in a [001]-oriented Ge/SiGe quantum well~\cite{hendrickxFastTwoqubitLogic2020,scappucciGermaniumQuantumInformation2020,hendrickxFourqubitGermaniumQuantum2021} with an applied static magnetic field ${\bf B}=(B_x,B_y,B_z)$. Following Ref.~\onlinecite{bulaevElectricDipoleSpin2007}, the effective Hamiltonian describing the pure HH hole whose effective spin is parallel or antiparallel to the magnetic field direction reads,
\begin{equation}
	\begin{aligned}
		H_{\rm{QD}}=\frac{\pi_x^2+\pi_y^2}{2m_{\parallel}}+U(x,y)+H_{\textrm{SO}}+\frac{1}{2} \bm{g} \mu_B \boldsymbol{B}\cdot \boldsymbol{\sigma},
	\end{aligned}
    \label{eq1}
\end{equation}
where $\boldsymbol{\pi}=\boldsymbol{p}+e\boldsymbol{A}$ is the usual Peierls substitution with the vector potential $\boldsymbol{A}$, $m_{\parallel}$ the in-plane HH effective mass,  $\bm{g}$ the Lande g-factor tensor of HH hole, $\boldsymbol{\sigma}$ the Pauli vector, and $\mu_B$ the Bohr magneton. The harmonic lateral  confining potential is $U(x,y)=\frac{1}{2}m_{\parallel}\omega_0^2 (x^2 + y^2)$, where $\omega_0=\hbar/m_{\parallel} r_0^2$ is the energy scale  that characterizes the lateral confinement for an effective QD lateral size $r_0$. Here, we take a gauge $\boldsymbol{A}=B_z(-y/2,x/2,0)$, considering negligible orbital effect induced by in-plane components $B_{x}$ and $B_{y}$  due to the strong quantization of motion along $z$~\cite{golovach2006electric,borhaniSpinManipulationRelaxation2012}.  In the absence of SOC ($H_{\textrm{SO}}=0$), we can label the eigenstates of Eq.~\eqref{eq1} as the product of Fock-Darwin and spin states $\vert n,l,s\rangle=\vert n,l\rangle \vert s\rangle$, where $n$, $l$ are the principle and azimuthal quantum numbers, respectively, and $s=\pm 3/2$. Fig.~\ref{fig1}(b) shows that each level is a Kramer's doublet that splits into two spin states in a magnetic field: $E_{n,l,s}=\hbar\Omega(n+1)+\hbar\omega_c l /2-\hbar\omega_Z s/3$ ($\Omega=\sqrt{\omega_0^2+\omega_c^2}$, $\omega_Z=\mu_B\bm{g}\cdot\bm{B}/\hbar$, and $\omega_c=eB_z/2m_{\parallel}$ is the cyclotron frequency). A qubit can be encoded into the lowest spin doublet. When applying an in-plane ac electric field $\boldsymbol{E_{ac}}(t)=E_{\textrm{AC}}(\textrm{sin}\omega t,\textrm{cos}\omega t,0)$ created by driving gates, electric-dipole transitions ($\Delta n=\pm1$, $\Delta s=0$) occur between the lowest  and higher excited doublets rather than within the lowest Zeeman-split spin doublet to yield spin-flip.

The situation might alter taking the SOC into account  ($H_{\textrm{SO}}\neq0$) since it entangles the orbitals with the different spins.  Because of the absence of bulk inversion asymmetry induced Dresselhaus SOC in centrosymmetric solids, structural inversion asymmetry induced Rashba effect (including interface effect) becomes the only source for SOC in Ge/Si QWs, in which a finite \textit{k}-linear term instead of (commonly thought) \textit{k}-cubic term has recently been recognized as the leading order in Rashba SOC of 2D HH holes~\cite{xiongEmergenceStrongTunable2021}.  This \textit{k}-linear Rashba SOC arises from a combination of local interface-induced HH-LH coupling and direct dipolar intersubband coupling to the external electric field. Since the \textit{k}-linear term tends to overwhelm all other higher-order terms that are very weak in Ge~\cite{winkler2003spin,philippNaturalHeavy-hole2021}, the effective SOC Hamiltonian reads,
\begin{equation}
	H_{\rm{SO}}=\frac{\alpha_{R}}{\hbar}(\pi_x\sigma_y-\pi_y\sigma_x),
\end{equation} 
where $\alpha_{R}$ is the Rashba parameter obtained from the atomistic pseudopotential calculations for Ge/Si QWs~\cite{xiongEmergenceStrongTunable2021}. Note that we only consider HH-LH coupling in obtaining $\alpha_{R}$. Taking $H_{\rm{SO}}$ into account as a perturbation to $H_{\textrm{QD}}$, we obtain the Zeeman-split ground spin doublet in the first-order perturbation theory as follows:
\begin{equation}
	\vert 0 \pm \rangle= \vert 0,0,\pm 3/2 \rangle+ \beta^{\pm}  \vert 1,\pm 1,\mp 3/2 \rangle, \label{eq2}
\end{equation} 
where $\beta^\pm = \pm \alpha_{R} m_{\parallel} \ell \omega_\pm / \hbar \omega^{\pm}_\alpha$, $\ell=\sqrt{\hbar/m_{\parallel} \Omega}$,  $\omega_\pm=\Omega \pm \omega_c$,  $\omega_\alpha^+=\omega_{+}+\omega_Z$, and $\omega_\alpha^-={\rm sgn}(\omega_- -\omega_Z)\sqrt{(\omega_- -\omega_Z)^2+[2\alpha_{R} m_{\parallel} \ell\omega_-/\hbar]^2}$. From Eq.~\eqref{eq2} we learn that electric-dipole transitions between $ \vert 0,0, -3/2 \rangle$ and $\vert 1,\pm 1,-3/2 \rangle$ will bring  $\vert 0 -\rangle$ (spin-down state $\vert\Downarrow \rangle $) to $\vert 0 + \rangle$ (spin-up state $\vert\Uparrow \rangle$), as shown in Fig.~\ref{fig1}(b). When the frequency of the electric field matches the spin resonance frequency of the qubit, stable Rabi oscillation occurs~\cite{bulaevElectricDipoleSpin2007}.

We now turn to estimate the Rabi frequency based on the effective Hamiltonians described above, following the procedure for the electron counterpart~\cite{borhaniSpinDecayQuantum2006,borhaniSpinManipulationRelaxation2012}. We refer the reader to supplementary materials (SM)~\cite{SM} for the details. Considering the Rabi frequency is strongly dependent on the static magnetic and ac electric field directions,  we study the case under in-plane and out-of-plane magnetic fields, separately. For the case under an in-plane magnetic field ${\bf B}= (B_x, 0, 0)$ (we set the ac electric field along the $x$-direction for simplicity), the Rabi frequency is as follows:
\begin{equation}
	f_{R}^{B_{\parallel}}=\frac{e E_{\textrm{AC}} \alpha_R g_\parallel \mu_BB_x}{2 \hbar\left(\hbar^{2} \omega_0^{2}- g_\parallel^2 \mu_B^2B_x^2\right)}.\label{eq3}
\end{equation}
For ${\bf B}= (0, 0, B_z)$, we obtain Rabi frequency
\begin{equation}
	f_{R}^{B_{\perp}}=\frac{e E_{\textrm{AC}} \alpha_R g_\perp \mu_BB_z}{2 \hbar\left(\hbar\omega_- +g_\perp \mu_BB_z\right)\left(\hbar\omega_+-g_\perp \mu_BB_z\right)}.
\end{equation}
The g factor is highly anisotropic in Ge: the in-plane g factor is $g_\parallel\sim 0.3$ measured in single-hole qubit experiments \cite{hendrickxSingleholeSpinQubit2020,hendrickxFastTwoqubitLogic2020}; nevertheless, the out-of-plane g factor is $g_\perp=15.7$ \cite{miller2021effective}. To have the same Zeeman splitting and thus Rabi frequency, a much smaller  magnetic field magnitude is required for the out-of-plane scenario than the in-plane scenario.

\begin{figure}[b] 
	\centering 
	\includegraphics[width=0.8\linewidth]{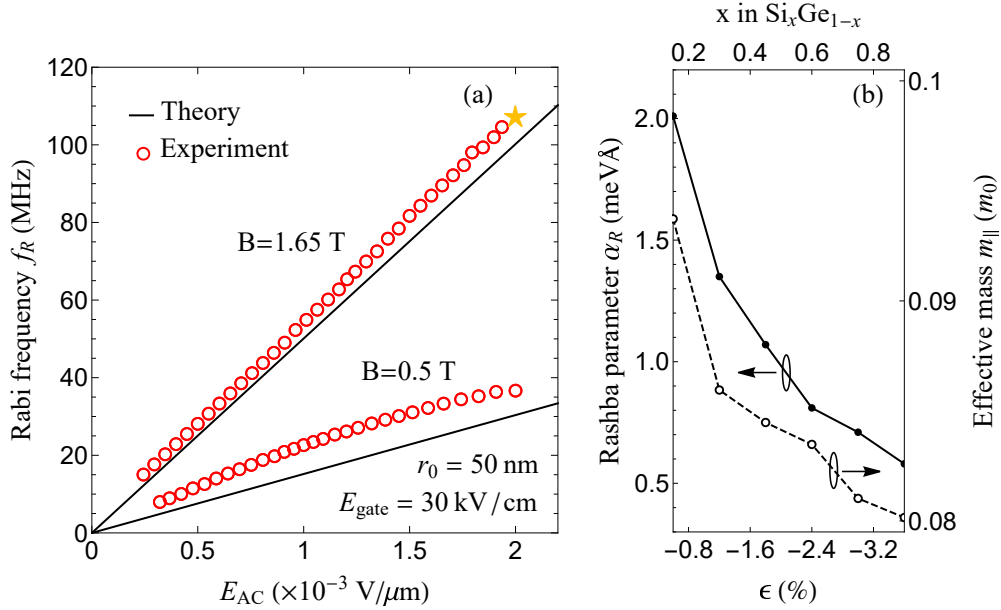} 
	\caption{(a)  The theoretically predicted Rabi frequency as a function of the amplitude of the ac electric field upon application of in-plane magnetic fields of $B=1.65$ T and $B=0.5$ T, respectively, compared to corresponding experimental results~\cite{hendrickxFastTwoqubitLogic2020} for planar Ge QD formed in a strained Ge/Si$_{0.2}$Ge$_{0.8}$ QW with Ge layer thickness of $L=16.7$ nm. We take an effective dot radius $r_0=50~\rm{nm}$ for $40-60$ nm dot lateral size in the experiment~\cite{PhysRevB.100.041304,sammakShallowUndopedGermanium2019}. For comparison, we have mapped the experimental microwave power given in Ref.~\onlinecite{hendrickxFastTwoqubitLogic2020} to $E_{\rm AC}$ regarding the maximum microwave power was estimated corresponding to $E_{\rm AC}=2\times 10^{-3}$ V/$\mu$m~\cite{boscoSqueezedHoleSpin2021}.  (b) The predicted $k$-linear Rashba parameter $\alpha_R$ and HH in-plane effective mass $m_{||}$ of strained Ge QW as a function of in-plane strain $\epsilon$ by carrying out atomistic pseudopotential method calculations. Because the strain of the Ge layer is induced by the Si$_x$Ge$_{1-x}$ alloy barrier, we can relate $\epsilon$ to the Si content $x$ in the $\rm{Si_xGe_{1-x}}$ alloy barrier by Vegard's law $\epsilon=-0.04 x$ \cite{Wortman_1965}. Here, we theoretically obtain $\epsilon=-0.8\%$ for Si$_{0.2}$Ge$_{0.8}$ alloy barrier, but we still adopt experimentally determined $\epsilon=-0.63\%$~\cite{sammakShallowUndopedGermanium2019} for calculations of $\alpha_R$ and $m_{||}$.} 
	\label{fig2} 
\end{figure}

In the following, we focus  on ${\bf B}= (B_x, 0, 0)$ scenario that has been adopted in recent experiments with achieved Rabi frequency exceeding 100 MHz for hole spin confined in gate-defined QDs~\cite{hendrickxSingleholeSpinQubit2020,hendrickxFastTwoqubitLogic2020,hendrickxFourqubitGermaniumQuantum2021} in strained Ge/Si$_{0.2}$Ge$_{0.8}$ QW with a 16 nm thick Ge layer~\cite{sammakShallowUndopedGermanium2019}. Fig.~\ref{fig2} (a) shows the calculated $f_{R}^{B_{\parallel}}$ as a function of driving amplitude $E_{\textrm{AC}}$ for $B=1.65$ T and $B=0.5$ T, respectively. To make a direct quantitative comparison with experimental results, here we calculate Rabi frequency $f_{R}^{B_{\parallel}}$ according to Eq.~\eqref{eq3} by employing experimental parameters (taken $r_0=50$ nm for dot lateral size in the range of 40-60 nm~\cite{hendrickxSingleholeSpinQubit2020}, $m_{\parallel}=0.09  m_0$ for Ge under an in-plane compressive strain of 0.63\%~\cite{sammakShallowUndopedGermanium2019}) except for the Rashba parameter $\alpha_{R}$. As shown in Fig.~\ref{fig2} (b), we predict $\alpha_{R}=2.01$ meV{\AA} by performing atomistic calculations for the corresponding Ge QW under an estimated biased electric field of 30 kV/cm for the experimental gate voltages~\cite{boscoSqueezedHoleSpin2021}.  One can see from Fig.~\ref{fig2} (a) that theoretically predicted Rabi frequency is in excellent agreement with experimental results over a wide range of driving amplitude for both magnetic fields. Specifically, the fastest Rabi frequency of 108 MHz~\cite{hendrickxFastTwoqubitLogic2020} was reached experimentally at $B=1.65$ T and $E_{\textrm{AC}} \approx 2 \times 10^{-3}$ V/$\mu$m~\cite{boscoSqueezedHoleSpin2021}, under which our theoretically predicted value is 100 MHz. The high agreement illustrates that the emergent \textit{k}-linear Rashba SOC via EDSR provides the fast hole spin control in planar Ge QDs.

\begin{figure}[b]                
	\centering 
	\includegraphics[width=0.8\linewidth]{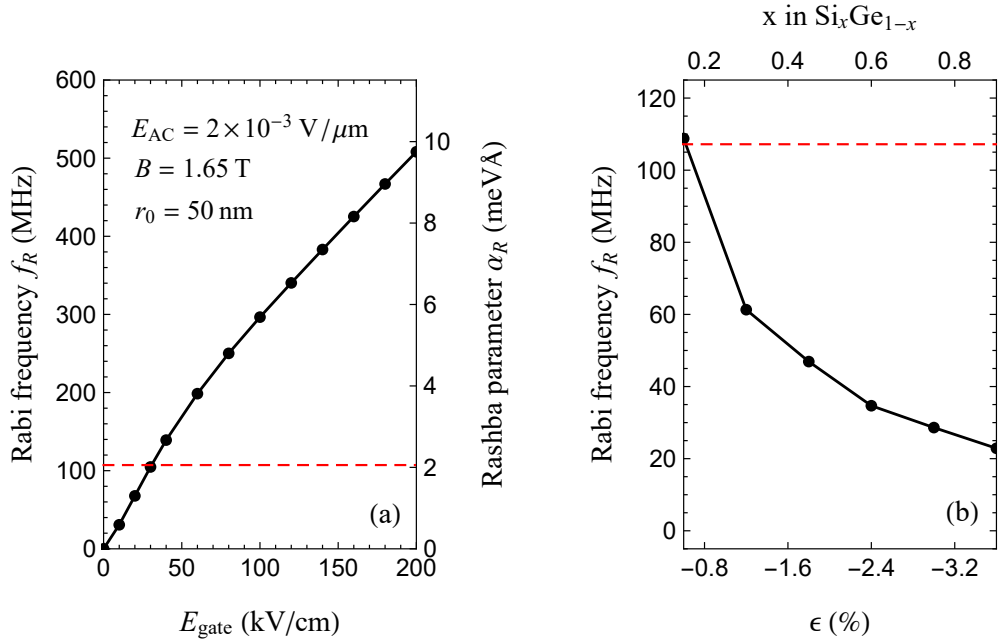} 
	\caption{The predicted Rabi frequency (and \textit{k}-linear Rashba parameter $\alpha_R$) as a function of gate electric field for $r_0=50~\rm{nm}$, $B=1.65$ T, and $E_{\rm AC}=2\times 10^{-3}~\rm{V/\mu m}$. The red dashed line indicates the maximum Rabi frequency of 108 MHz achieved experimentally~\cite{hendrickxFastTwoqubitLogic2020}.  (b) The corresponding Rabi frequency as a function of  biaxial strain in the Ge layer.} 
	\label{fig3} 
\end{figure} 

Fig.~\ref{fig2}(b) shows that raising the Si content $x$ in the $\rm{Si_{x}Ge_{1-x}}$ barrier will linearly enhance the compressive strain in the Ge layer~\cite{sammakShallowUndopedGermanium2019,Wortman_1965} because Si has a lattice constant 4.3\% smaller than that of Ge. One can see that the enhanced compressive strain, in turn, causes a reduction in both Rashba parameter $\alpha_R$ and in-plane HH effective mass $m_{\parallel}$, which is consistent with the experimental observations~\cite{Sawano_2011}. The reduction in $m_{\parallel}$ will benefit the enhancement of hole mobility. However, it also yields a detriment in the Rabi frequency combining with the reduction of $\alpha_R$, as shown in Fig.~\ref{fig3}(b). Hence, a low Si content in the $\rm{Si_{x}Ge_{1-x}}$ barrier is demanded to achieve a high Rabi frequency.

The strength of the Rashba SOC is usually electrically tunable by biased gates, which provides a feasible way to enhance the Rabi frequency further. We examine the Rabi frequency by varying gate electric field $E_{gate}$ applied to Ge/Si$_{0.2}$Ge$_{0.8}$ QW whose $\alpha_R$ is obtained from the atomistic calculations.  Fig.~\ref{fig3}(a) exhibits that Rabi frequency grows up linearly as we amplify the gate electric field due to the enhancement of Rashba SOC strength $\alpha_R$.  The Rabi frequency boosts to 500 MHz at a 200 kV/cm gate electric field compared with the reported 108 MHz at 30 kV/cm~\cite{hendrickxFastTwoqubitLogic2020}. 

So far, we have demonstrated that the emergent \textit{k}-linear Rashba SOC drives the fast Rabi frequency achieved experimentally. However, this \textit{k}-linear Rashba SOC is relatively week ($\alpha_R<10$ meV\AA)  in [001]-oriented Ge QWs compared with [110]-oriented counterparts where $\alpha_R$ exceeds 120 meV\AA~\cite{xiong2021upper} and Ge nanowires where $\alpha_R$ is predicted over 400 meV\AA~\cite{luoRapidTransitionHole2017}. We thus expect that Rabi frequency can reach multiple GHz for gate-defined QDs formed in [110]-oriented Ge/Si QWs and Ge nanowires. In addition, the asymmetry in the lateral confinement potential of QDs has also been suggested to enhance Rabi frequency~\cite{boscoSqueezedHoleSpin2021}. Furthermore, Terrazos \textit{et al.}~\cite{terrazosTheoryHoleSpinQubits2021} recently invoked a cubic-symmetric component of the cubic-in-\textit{k} Rashba SOC to explain the experimentally achieved Rabi frequency by assuming an extremely large cubic Rashba parameter $6\times 10^4$ meV\AA$^3$~\cite{Wang_2021}. However, our atomistic calculation predicts that the total cubic Rashba parameter is less than $10^3$ meV\AA$^3$ in the Ge QW under the same electric field of 100 kV/cm as Ref~\cite{Wang_2021}, and the contribution of the cubic-symmetric component to the total cubic Rashba parameter is tiny in Ge since it is proportional to $(\gamma_2-\gamma_3)/2$~\cite{philippNaturalHeavy-hole2021}. Besides SOC-driven EDSR, there is another mechanism contributing to EDSR. It is known as g-tensor magnetic resonance (g-TMR), which utilizes the gate-voltage modulation of a g-matrix~\cite{PhysRevLett.120.137702, Kato1201}. Crippa \textit{et al.} have discriminated the contributions of these two mechanisms to Rabi frequency for Si hole spin qubit in the nanowire and found the SOC mechanism to be the main contributor to the Rabi frequency~\cite{PhysRevLett.120.137702}. Specifically, the g-TMR mechanism is negligible when the magnetic field is applied in-plane along the nanowire direction~\cite{PhysRevLett.120.137702} in the same configure as investigated here. There are other studies that have investigated the microscopic mechanism of hole spin-orbit coupling in Ge/Si QWs.

In conclusion, we have resolved the standing puzzle by identifying the emergent \textit{k}-linear Rashba SOC in 2D holes as the driver via EDSR for the rapid hole spin manipulation achieved experimentally. Because the \textit{k}-linear Rashba SOC is electrically tunable, we suggest using the applied gate electric field to enhance the Rabi frequency exceeding 500 MHz. We can further boost the Rabi frequency to multiple if we replace [001]-oriented Ge QW by [110]-oriented counterpart.

%\nocite{*}
%\bibliographystyle{apsrev4-2}
\bibliography{arxiv.bib}% Produces the bibliography via BibTeX.

\end{document}